\documentclass[letter,fleqn,usenatbib]{mnras}

\usepackage{newtxtext,newtxmath}
\usepackage[T1]{fontenc}

\DeclareRobustCommand{\VAN}[3]{#2}
\let\VANthebibliography\thebibliography
\def\thebibliography{\DeclareRobustCommand{\VAN}[3]{##3}\VANthebibliography}

\usepackage{graphicx}	
\usepackage{amsmath}	
\usepackage{array}

\newcommand{\ie}{i.e.,~}
\newcommand{\eg}{e.g.,~}

\title[Impact of mass constraints on neutron stars]{Impact of large-mass constraints on the properties of neutron stars}

\author[C. Ecker and L. Rezzolla]{
Christian Ecker$^{1}$\thanks{E-mail: ecker@itp.uni-frankfurt.de (CE)}
and Luciano Rezzolla$^{1,2,3}$
\\
$^{1}$Institut f\"ur Theoretische Physik, Goethe Universit\"at, Max-von-Laue-Str. 1, 60438 Frankfurt am Main, Germany\\
$^{2}$School of Mathematics, Trinity College, Dublin 2, Ireland\\
$^{3}$Frankfurt Institute for Advanced Studies,
  Ruth-Moufang-Str. 1, 60438 Frankfurt am Main, Germany
}

\date{Accepted XXX. Received YYY; in original form ZZZ}
\pubyear{2015}

\begin{document}
\label{firstpage}
\pagerange{\pageref{firstpage}--\pageref{lastpage}}
\maketitle

\begin{abstract}
The maximum mass of a nonrotating neutron star, $M_{\rm TOV}$, plays a
very important role in deciphering the structure and composition of
neutron stars and in revealing the equation of state (EOS) of nuclear
matter. Although with a large-error bar, the recent mass estimate for the
black-widow binary pulsar PSR J0952-0607, \ie $M=2.35\pm0.17~M_\odot$,
provides the strongest lower bound on $M_{\rm TOV}$ and suggests that
neutron stars with very large masses can in principle be
observed. Adopting an agnostic modelling of the EOS, we study the impact
that large masses have on the neutron-star properties. In particular, we
show that assuming $M_{\rm TOV}\gtrsim 2.35\,M_\odot$ constrains tightly
the behaviour of the pressure as a function of the energy density and
moves the lower bounds for the stellar radii to values that are
significantly larger than those constrained by the NICER measurements,
rendering the latter ineffective in constraining the EOS. We also provide
updated analytic expressions for the lower bound on the binary tidal
deformability in terms of the chirp mass and show how larger bounds on
$M_{\rm TOV}$ lead to tighter constraints for this quantity. In addition,
we point out a novel quasi-universal relation for the pressure profile
inside neutron stars that is only weakly dependent from the EOS and the
maximum-mass constraint. Finally, we study how the sound speed and the
conformal anomaly are distributed inside neutron stars and show how these
quantities depend on the imposed maximum-mass constraints.
\end{abstract}


\section{Introduction}

The maximum mass beyond which a static relativistic star collapses to a
black hole, $M_{\rm TOV}$, is determined by the solution of the
equilibrium equations for a self-gravitating fluid configuration -- the
so-called Tolmann-Oppenheimer-Volkoff (TOV) equations -- once an equation
of state (EOS), that is a relation between the pressure and the energy
density $p(e)$, is specified. Given the intimate relation between the EOS
and the maximum mass, the knowledge of the latter has always been
considered an essential tool to access the former. 

Chiral Effective Theory (CET) calculations~\citep{Hebeler:2013nza,
  Gandolfi2019, Keller:2020qhx, Drischler:2020yad} constrain the EOS at
baryon densities $n$ below and around nuclear saturation density
$n_s=0.16\,{\rm fm}^{-3}$. At densities much higher than those realised
inside neutron stars ($n\gg \,n_s$), matter is in a state of deconfined
quarks and gluons and the EOS of Quantum Chromodynamics (QCD) becomes
accessible to perturbation theory~\citep{Freedman:1976ub,
  Vuorinen:2003fs, Gorda:2021kme}. Between these limits, at densities a
few times larger than $n_s$, such as those realised in neutron-star
cores, these methods are not applicable, hence our knowledge about even
the most basic neutron-star properties like their mass-radius relation
and in particular their maximum mass is incomplete. In this regime the
currently available theoretical options are specific-model
building~\citep[see, \eg][for some recent works]{Bastian:2020unt,
  Demircik:2021zll, Ivanytskyi:2022wln}, and model agnostic
EOS-samplings~\citep[see, \eg][for some recent attempts]{Greif2019,
  Annala2019, Dietrich:2020efo, Altiparmak:2022}, for which CET and QCD
provide important constraints~\citep{Komoltsev:2021jzg,Gorda:2022,Somasundaram:2022}.

In addition, a number of EOS independent quasi-universal relations have
been identified among various neutron-star properties, either when
isolated~\citep[see, \eg][]{Yagi2013a} or when in binary
systems~\citep[see, \eg][]{Baiotti2016}. These relations provide a useful
tool to break the degeneracy between existing uncertainties in the EOS
and differences between General Relativity and alternative theories of
gravity. Quasi-universal relations have also been found to describe the
critical mass of equilibrium models with varying angular momentum, that
is, the maximum mass along the stability line of uniformly rotating
configurations. In turn, this relation allows one to tightly constrain
the ratio between the maximum mass of uniformly rotating and static stars
made of purely nucleonic matter, \ie $M_{\rm
  max}=1.203^{+0.022}_{-0.022}\,M_{\rm TOV}$~\citep{Breu2016} and to only
slightly larger values when accounting for a phase transition to quark
matter~\citep{Bozzola2019, Demircik:2020jkc}.

On the observational side, direct mass measurements of
$M\approx2\,M_\odot$~\citep{Antoniadis:2013pzd, NANOGrav:2019jur,
  Fonseca:2021wxt}, combined with mass and radius measurements by the
NICER experiment~\citep{Riley:2019yda, Miller:2019cac, Miller2021,
  Riley:2021pdl} and with measurements of the binary tidal deformability
$\tilde\Lambda$ from the binary neutron-star merger
GW170817~\citep{Abbott2018a}, have provided until recently the most
important benchmarks for the EOS at densities beyond $n_s$. The direct
mass measurement of the heavy companion in the black-widow binary pulsar
PSR~J0952-0607, namely $M=2.35\pm0.17~M_\odot$ by~\citet{Romani:2022jhd},
significantly exceeds the results of any previous mass
measurements. Although reported with a large uncertainty, such a
measurement represents currently the most massive known neutron star and
provides the strictest lower bound on the maximum neutron-star mass to
this date. This \textit{observational} mass measurement, at least in its
lower bound, is still compatible with the theoretical predictions made
for the maximum mass by a number of groups on the basis of the GW170817
event and the corresponding gamma-ray burst event GRB170817A, namely
$M_{\rm TOV}\lesssim 2.16^{+0.17}_{-0.15}
\,M_{\odot}$~\citep{Margalit2017, Rezzolla2017, Ruiz2017,
  Shibata2019}. In addition, recent work taking into account the GW190814
event, has shown that maximum masses in excess of $\simeq 2.4\,M_{\odot}$
have problems satisfying the limits on the gravitational mass emitted in
gravitational waves or the rest-mass ejected after the
merger~\citep{Nathanail2021}.

Obviously, the novel mass constraint provided by PSR~J0952-0607, and the
future refinements that are expected to reduce the measurement
uncertainties, will provide very important input to further sharpen the
focus on the properties of neutron stars and, in particular, on some of
the most basic features of any EOS. One such quantity is the adiabatic
sound speed ~\citep{Rezzolla_book:2013}
\begin{equation}
c_s^2:=\left(\frac{ \partial  p}{ \partial  e}\right)_s\,,
\end{equation}
where $p, e$, and $s$ are the pressure, energy density and specific
entropy, respectiviely; clearly, the sound speed is bounded by causality
and thermodynamic stability to $0\leq c_s^2\leq 1$. Because the sound
speed provides a direct measure of the stiffness of matter within the
star, it represents a very accurate tool to probe the stellar
interior and is obviously directly related the maximum mass that any EOS
can support. The properties of the sound speed at finite densities have
been studied extensively with various approaches in the recent past
\citep[see, \eg][and references therein]{Ecker:2017fyh, McLerran:2018hbz,
  Leonhardt:2019fua, Margueron:2021dtx, Duarte:2021tsx, Pal:2021qav,
  Altiparmak:2022, Brandes:2022, Braun:2022, Ecker:2022}. There is now
widespread consensus that the sound speed is much smaller than the speed
of light ($c_s^2\ll 1$) in low-density matter ($n\lesssim n_s$) and
approaches the conformal limit $c_s^2=1/3$ from below at large densities
($n\gtrsim 40\,n_s$). In between, the sound speed is most likely
non-monotonic and reaches a local maximum that exceeds the conformal
value $c_s^2>1/3$ at densities few times larger than
$n_s$~\citep{Altiparmak:2022}. Because this local maximum in the sound
speed at densities typically probed by neutron-star interiors was already
necessary to explain the previous bounds on the maximum-mass measurements
$M\gtrsim2~M_\odot$~\citep[see the arguments by][]{Bedaque2015,
  Hoyos:2016cob, Moustakidis2017, Kanakis-Pegios:2020jnf}, the recent
bound set by PSR~J0952-0607 has the direct consequence of moving the
maximum to even larger values and lower densities.

Another quantity related to the stellar interior that has been recently
attracted considerable interest \citep[see, \eg][]{Fujimoto:2022,
  Marczenko:2022} is the so-called ``conformal anomaly'', that is, the
normalized trace of the star's energy-momentum tensor $T^{\mu\nu}$
\begin{equation}
\label{eq:Delta}
  \Delta:= \frac{1}{3}\frac{g_{\mu\nu}T^{\mu\nu}}{e} =
  \frac{1}{3}-\frac{p}{e}\,,
\end{equation}
where $g_{\mu\nu}$ is the metric tensor and the second equality in
Eq.~\eqref{eq:Delta} refers to a perfect fluid and is true for any metric
and coordinate system~\citep{Rezzolla_book:2013}. Requiring that the
conformal anomaly satisfies causality and thermodynamic stability leads
to the following allowed range for $\Delta$
\begin{equation}
-\frac{2}{3}\leq \Delta\leq \frac{1}{3}\,.
\end{equation}
The interest in this quantity stems from the fact that it provides a
simple measure of the deviation from conformal symmetry for
matter at nuclear and super-nuclear densities. We recall that in QCD the
coupling runs with energy, which introduces a scale that breaks conformal
symmetry at finite densities and/or temperature. Only at asymptotically
large temperatures and/or densities, conformal symmetry is restored and
$\Delta=0$. At finite densities, however, both the value and the sign of
the conformal anomaly are not know, although a conjecture was put forth
that $\Delta\geq 0$ in neutron-star
interiors~\citep{Fujimoto:2022}. Here, we test this conjecture and show
that besides depending on the lower bound imposed on the maximum mass,
the conformal anomaly can vary significantly both in size and sign.

Finally, improved mass measurements of pulsars also induce constraints on
the properties of binary neutron-star systems, that can be tested with
gravitational-wave detections. For example, the first ever detected
gravitational-wave signal of a binary neutron-star merger, \ie GW170817,
set an upper bound on the binary tidal deformability parameter
$\tilde\Lambda\leq720$~\citep{Abbott2018a}, while subsequent early
studies have further restricted such uncertainty~\citep[see,
  \eg][]{Radice2017b, Most2018}. Using the less conservative maximum-mass
bounds of $M_{\rm TOV} \gtrsim 2.0\,M_{\odot}$,~\citet{Altiparmak:2022}
have recently proposed a simple analytic expression to determine the
upper and lower bounds of the tidal deformability in terms of the chirp
mass of the binary, \ie $\tilde\Lambda_{\rm min/max}(\mathcal{M}_{\rm
  chirp})$. We here reconsider these expressions in the light of the more
extreme bounds set by the mass measurement in PSR~J0952-0607 and show
that for any value of $\mathcal{M}_{\rm chirp}$, only $\tilde\Lambda_{\rm
  min}$ depends on the assumed maximum mass, while $\tilde\Lambda_{\rm
  max}$ remains essentially unchanged with respect to the expression
presented by~\citet{Altiparmak:2022}.


\section{Methods}
\begin{figure*}
    \centering
    \includegraphics[width=0.475\textwidth]{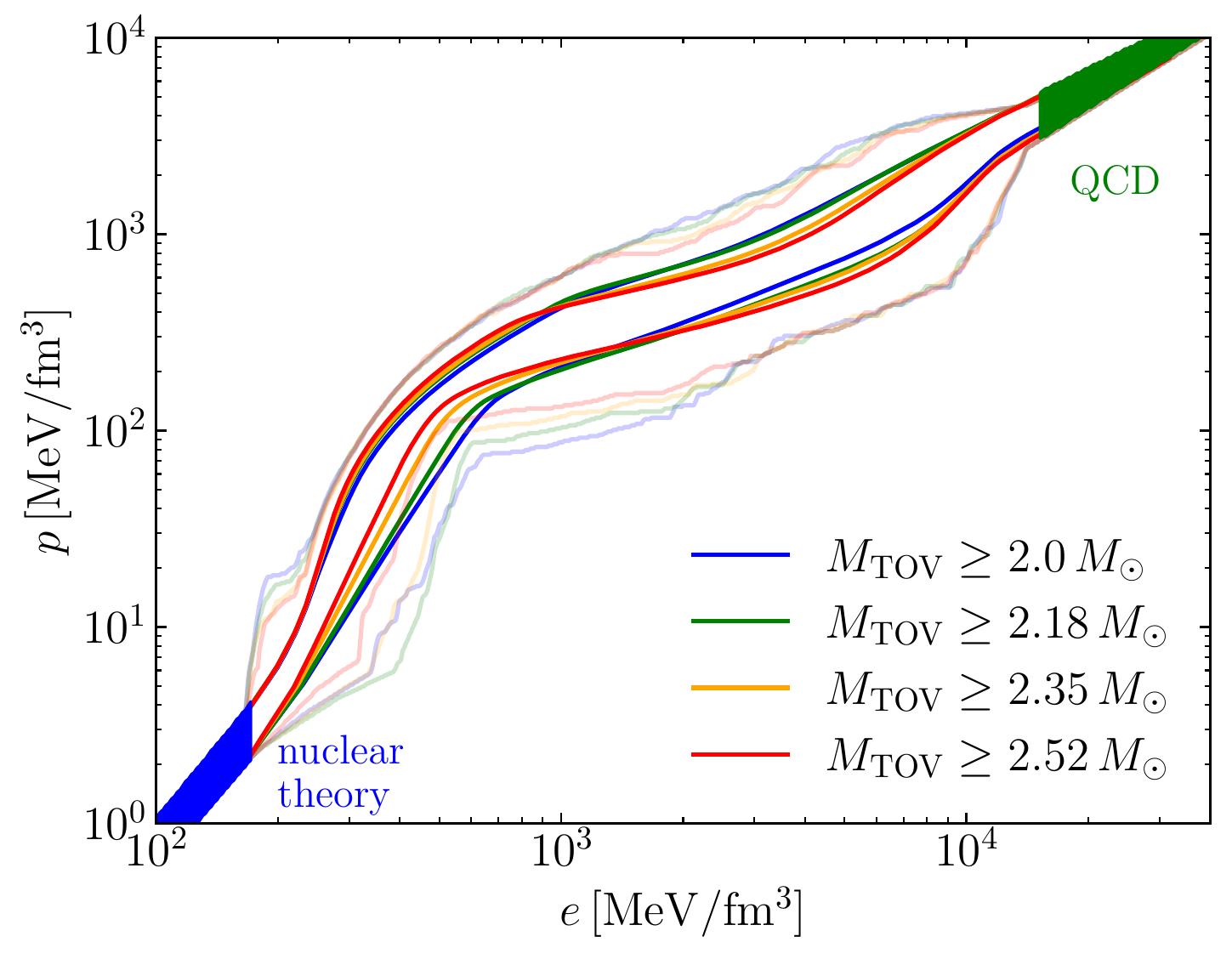}\quad
    \includegraphics[width=0.475\textwidth]{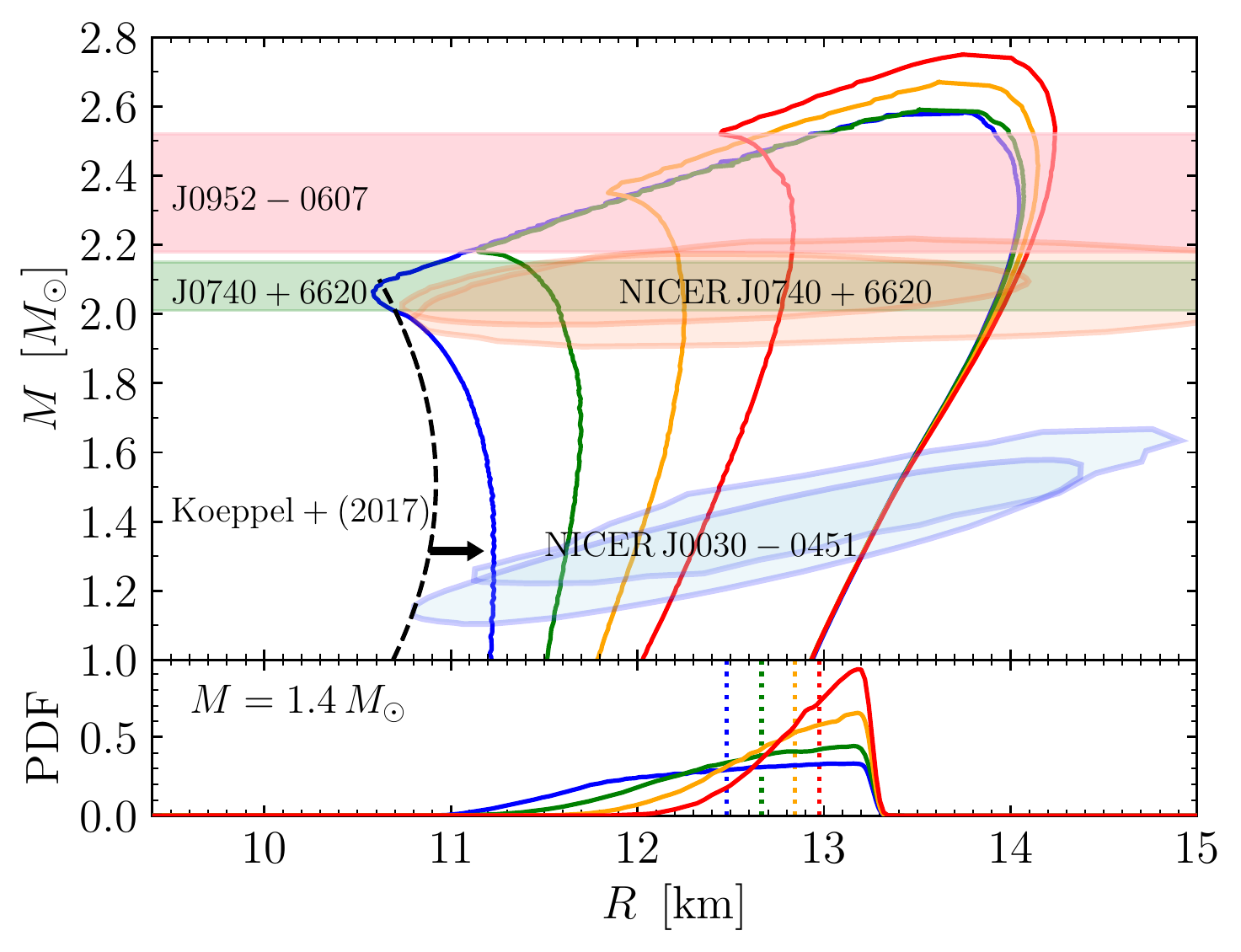}\quad
    \caption{\textit{Left Panel:} PDFs of the various EOSs constructed,
      with coloured lines showing the $95\%$-confidence intervals for the
      different mass constraints. Lines with light colours indicate
      instead the outer boundaries to excluded regions. The blue and
      green-shaded areas mark the uncertainty of nuclear theory and
      perturbative QCD, respectively. \textit{Right Panel:} The same as
      in the left panel but for the PDFs of the mass-radius
      relations. Blue and orange ellipses are radius measurements of
      J0030+0451~\citep{Riley:2019yda, Miller:2019cac} and of
      J0740+6620~\citep{Miller2021, Riley:2021pdl} by the NICER
      experiment, respectively. Green and pink areas are mass
      measurements of J0740+6620~\citep{Fonseca:2021wxt} and
      J0952-0607~\citep{Romani:2022jhd}, respectively. In the bottom part
      of the panel is reported slices of the PDF for $M=1.4\,M_\odot$,
      with the medians being marked by vertical dotted lines.}
    \label{fig:EOS-MR_contour}
\end{figure*}

To construct the EOSs we employ in our analysis, we follow the procedure
by~\citet{Altiparmak:2022} and \citet{Ecker:2022}, to which we refer
for more details. In essence, for densities below $n=0.5\,n_s$ we
use the Baym-Pethick-Sutherland (BPS) model~\citep{Baym71b} and extend it
until $n=1.2\,n_s$ with random polytropes bounded by the soft and stiff
EOSs of~\citet{Hebeler:2013nza}. For densities $n>1.2\,n_s$, we use the
sound-speed interpolation method introduced by~\citet{Annala2019} and
impose at $n\gtrsim 40\,n_s$ the parametrized next-to-next-to leading
order (2NLO) perturbative QCD results of~\citep{Fraga2014} (see also
\citet{Gorda:2021znl, Gorda:2021kme} for partial 3NLO improvement
and~\citet{Komoltsev:2021jzg} on how to propagate these constraints down
to neutron-star densities); we collectively refer to these as to the
``QCD'' constraints. Note that in contrast to~\citet{Altiparmak:2022} and
\citet{Ecker:2022}, we now also perform separate simulations where we
do not impose the QCD constraint so as to check its impact on the
results.

We also impose constraints from pulsar measurements and binary
neutron-star merger observations, to which we collectively refer as
``astro'' constraints. In particular, we impose the constraints deriving
from the radius measurements by the NICER experiment of
J0740+6620~\citep{Miller2021,Riley:2021pdl} and of
J0030+0451~\citep{Riley:2019yda, Miller:2019cac} by rejecting EOSs with
$R<10.75\,{\rm km}$ at $M=2.0\,M_\odot$ and $R<10.8\,{\rm km}$ at
$M=1.1\,M_\odot$, respectively (see Fig.~\ref{fig:EOS-MR_contour}). In
addition, we impose the upper bound on the binary tidal deformability
parameter $\tilde \Lambda \leq 720$ (low-spin priors) obtained from
GW170817~\citep{Abbott2018a}. Denoting respectively with $M_{i}$,
$R_{i}$, and $\Lambda_{i}$ the masses, radii, and tidal deformabilities
of the binary components, where $\Lambda_i :=
\tfrac{2}{3}k_2\left({R_i}/{M_i}\right)^5$, $i=1,2$, and $k_2$ is the
second tidal Love number, we compute the binary tidal deformability as
\begin{equation}
\tilde{\Lambda}:=\frac{16}{13}
\frac{\left(12M_2+M_1\right)M_{1}^4\Lambda_1+\left(12M_1+M_2\right)M_{2}^4\Lambda_2}
     {\left(M_1+M_2\right)^5}\,.
\end{equation}
For any choice of $M_{1,2}$ and $R_{1,2}$, we then reject those EOSs with
$\tilde\Lambda>720$ for a chirp mass $\mathcal{M}_{\rm chirp}:=(M_1
M_2)^{3/5}(M_1+M_2)^{-1/5}=1.186 M_{\odot}$ and $q:={M_2}/{M_1}>0.73$ as
required for consistency with LIGO/Virgo data for
GW170817~\citep{Abbott2018a}.

Because the limits on the maximum mass are still rather uncertain, we
perform separate simulations imposing different lower limits on the
maximum mass, namely, we consider $M_{\rm TOV}\geq
2.0,2.18,2.35,2.52~M_\odot$, where the first value is motivated by the pulsars
PSR~J0348+0432~\citep{Antoniadis:2013pzd} ($M=2.01\pm0.04~M_\odot$) and
PSR~J0740+6620~\citep{NANOGrav:2019jur,Fonseca:2021wxt}
($2.08\pm0.07~M_\odot$), while the last three values correspond to the
lower bound, the median and the upper bound of the uncertainty in the mass estimate
reported by~\citet{Romani:2022jhd} for PSR~J0952-0607
($2.35\pm0.17~M_\odot$). Note that~\citet{Romani:2022jhd} also reports a
more conservative estimate of $M_{\rm TOV} > 2.09\,M_\odot$, which we also checked;
however because the results are almost indistinguishable from the case
$M_{\rm TOV}\geq 2.0\,M_\odot$, we do not discuss them here. For each
mass-bound considered, we have constructed $\approx 10^6$ different
neutron star solutions passing all the QCD and astro constraints.

We remark that in our approach a certain bias inherited from the way the
prior is constructed is unavoidable. In our previous
works~\citep{Altiparmak:2022, Ecker:2022} we first sampled a temporary
maximum value for the maximally allowed sound speed $c^2_{s,{\rm max}}$
uniformly in the interval $[0,1]$ and then the various sound-speed values
at the individual matching points for the construction of the EOS on the
range $[0,c^2_{s,{\rm max}}]$. This approach guaranteed a sufficient
sampling rate for globally monotonic and sub-conformal EOS families,
which otherwise would be statistically suppressed.  Because here we are not
interested in such particular subsets of EOSs, we can omit the first step
and simply sample the individual sound-speed values between zero and one
directly.  This approach results in slightly higher estimates for the
sound speed maxima and also slightly different estimates for neutron-star
radii compared to our previous work.  The resulting difference for the
neutrons star radii $R_{1.4(2.0)}$ is negligible, being less than $40
(200)~\rm m$, however, larger differences can appear in the PDF of the
maximum sound-speed. One way to mitigate this intrinsic and inevitable
bias introduced by the choice of sampling is to employ a Bayesian
analysis. A comparison work between our approach and a fully Bayesian
approach is presently in progress and will be presented in an upcoming
work~\citep{Jiang2022}.

Before turning to the results, an important remark is worth making. With
a spin frequency of ${f=706\,\rm Hz}$~\citep{Romani:2022jhd}
PSR~J0952-0607 is the second-fastest-spinning pulsar known. This raises
the question of whether the static approximation assumed in our analysis
is actually justified and if it is not instead necessary to introduce
rotation-induced corrections. To address this question it is sufficient
to consider the approximate but analytic quasi-universal expression for
the critical mass along the dynamical stability line to gravitational
collapse, that is, the value of the maximum mass of a uniformly rotating
star when expressed as a function of the dimensionless angular momentum
$j:=J/M^2$~\citep{Breu2016}
\begin{equation}\label{eq:Breu}
	\frac{M_{\rm crit}}{M_{\rm TOV}}=1+a_2\,\left(\frac{j}{j_{\rm
            Kep}}\right)^2+a_4\,\left(\frac{j}{j_{\rm Kep}}\right)^4\,,
\end{equation}
where $j_{\rm Kep}$ is the Keplerian dimensionless angular momentum
($j/j_{\rm Kep}\leq1$) and the coefficients have values $a_2=0.1316$,
$a_4=0.07111$. Assuming a Keplerian frequency for PSR~J0952-0607 of
$f_{\rm Kep}\approx 1.5\,\rm kHz$ (see Table~1
of~\citet{Demircik:2020jkc}) and expression~\eqref{eq:Breu}, it is
possible to deduce that, in the case of PSR J0952-0607, ${f}/{f_{\rm
    Kep}}\approx{j}/{j_{\rm Kep}}\lesssim 0.46$, so that the
corresponding increase in the maximum mass is less than $0.1\,M_\odot$,
or, equivalently, less than $4\%$~\citep[see also Fig.~4
  of][]{Demircik:2020jkc}. In other words, given the much larger
uncertainties affecting the maximum mass, the use of the static
approximation is well justified and has no relevant impact on our
results.

\section{Results}

We first show in the left panel of Fig.~\ref{fig:EOS-MR_contour} the
$95\%$-confidence intervals (coloured lines) for the EOSs assuming
different maximum-mass bounds, together with the corresponding outer
envelopes (light coloured lines), where the blue and green shaded areas
mark theoretical uncertainties of nuclear theory and perturbative QCD,
respectively. Note how larger values for the bound on $M_{\rm TOV}$ lead
to a steeper increase of the pressure at energy densities below $e\approx
1\,{\rm GeV/fm}^3$. This clearly is a consequence of the increased
stiffness necessary to satisfy the larger bounds on $M_{\rm TOV}$ at low
densities. The effect is most prominent in the range $0.5-1\,{\rm
  GeV/fm}^3$ and pushes the lower $95\%$ contour of the pressure to
significantly larger values, while leaving the upper contour almost
unchanged. Thus, a simple and interesting result follows from this
analysis: large values of $M_{\rm TOV}$ tightly constrain the pressure to
be $p\approx 200\,{\rm MeV/fm}^3$ at $e\approx 600\,{\rm MeV/fm}^3$. On
the other hand, in the range $e\approx 1-10\,{\rm GeV/fm}^3$ the
increased mass constraint has the opposite effect: larger bounds on
$M_{\rm TOV}$ force the pressure to rise less rapidly.  Interestingly, in
the intermediate region, (\ie at $e\approx 1\,{\rm GeV/fm}^3$), the bound
on $M_{\rm TOV}$ has almost no impact and the EOSs are entirely
insensitive to the mass constraints.
 
In the right panel of Fig.~\ref{fig:EOS-MR_contour} we show instead the
$95\%$-confidence intervals of the mass-radius relations for different
maximum-mass bounds (coloured lines). Clearly, large-mass constraints lead
to a significant exclusion of stars with small radii, while the upper
limit at large radii ($R\approx13-14\,\rm km$) remains unaffected. The
systematic shift towards stellar models with larger radii is clearly a
consequence of the stiffening of the EOSs at $e\lesssim 1\,{\rm
  GeV/fm}^3$ shown in the left panel; without this additional pressure
support, it is not possible to construct equilibrium models with such
large maximum masses. In turn, this means that causality and QCD
together with the constraint $M_{\rm TOV} \gtrsim 2.18\,M_\odot$,
completely dominate the lower bounds for the stellar radii, rendering the
existing radius measurements by the NICER experiments -- and which only
provide lower limits for the radii -- essentially ineffective. The bottom
part of the right panel of Fig.~\ref{fig:EOS-MR_contour} reports with
coloured lines slices of the probability density functions (PDF) and the
corresponding median estimates (dotted lines) for the radii of a typical
neutron star with a mass of $M=1.4\,M_\odot$ (see also
Table~\ref{tab:table2} in Appendix\ref{sec:appendix}). Larger values for
$M_{\rm TOV}$ result in PDFs with significantly smaller probability at
small radii, while the sharp edge at large radii set by the tidal
deformability constraint remains essentially unaffected. As a result, the
median values are shifted to larger values and the uncertainties become
smaller, which can be seen more explicitly from the numbers provided in
Table~\ref{tab:table2} of Appendix~\ref{sec:appendix}.

\begin{figure}
    \centering
    \includegraphics[width=0.475\textwidth]{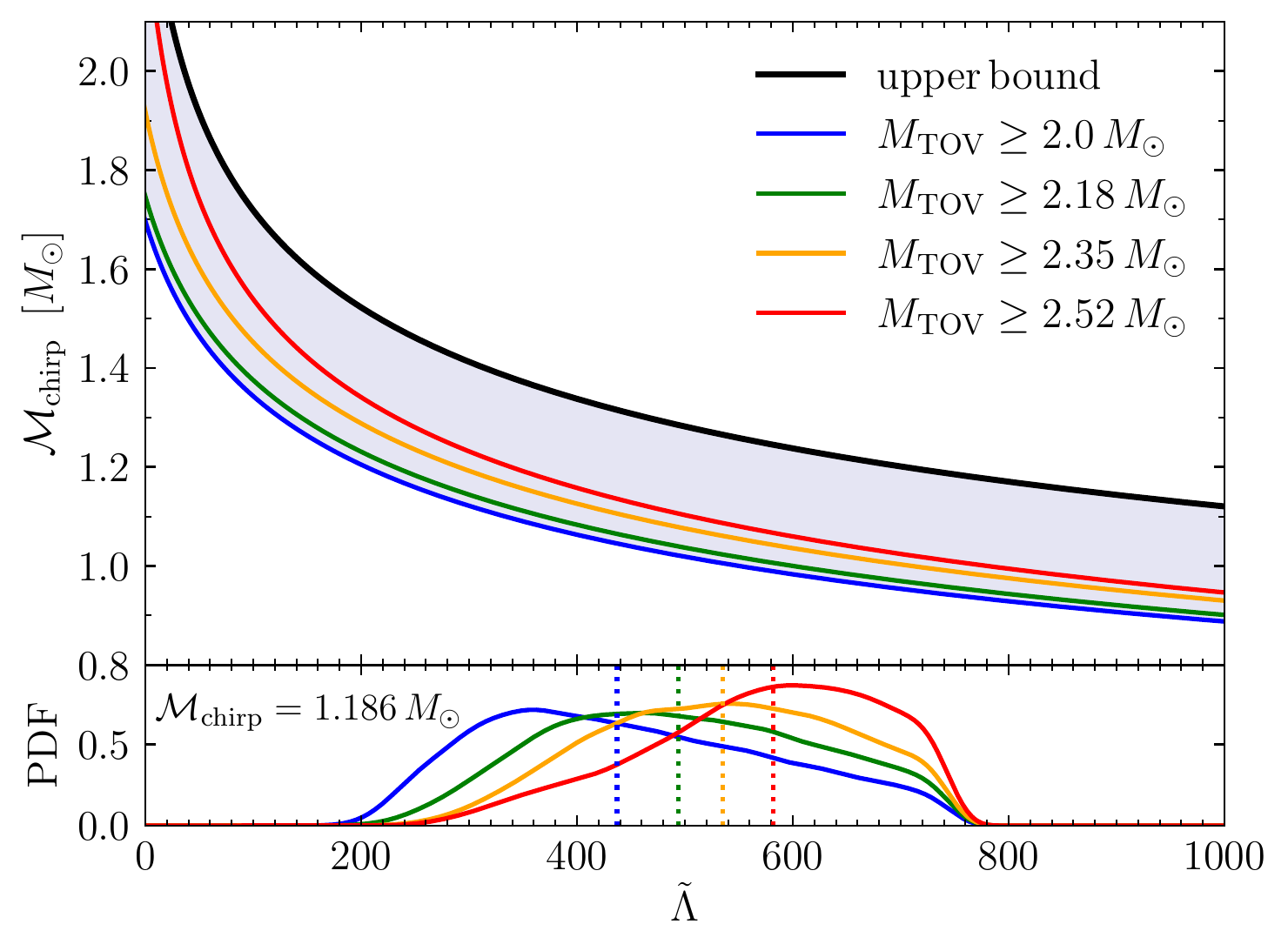}
    \caption{Relation between the chirp mass and binary tidal
      deformability. Coloured lines mark lower bounds of the
      $95\%$-confidence intervals for $\tilde{\Lambda}_{\rm min}$, while
      the black line is the upper bound $\tilde{\Lambda}_{\rm max}$,
      which is valid for all mass constraints. In the bottom part of the
      panel are reported the PDF slices for the measured chirp mass of
      GW170817 $\mathcal{M}_{\rm chirp}=1.186\,M_\odot$, while the
      medians are again marked with vertical dotted lines.}
    \label{fig:Lambda}
\end{figure}

Figure~\ref{fig:Lambda} is used to highlight the impact that a larger
mass bound has on the behaviour of the binary tidal deformability. In
particular, the figure shows the upper and lower bounds for $\tilde
\Lambda$ as function of the chirp mass
$\mathcal{M}_{\rm chirp}$. Such bounds were first presented
by~\citet{Altiparmak:2022}, where they were shown to follow the simple
relation
\begin{equation}\label{eq:Lambda}
    \tilde\Lambda_{\rm min (max)}=a+b\,\mathcal{M}_{\rm chirp}^c\,.
\end{equation}
This is a particularly important result, since it provides theoretical
predictions for the upper and lower bounds on $\tilde\Lambda$, a quantity
that constrains the EOSs, from $\mathcal{M}_{\rm chirp}$, a quantity that
can be (and has been) measured to high accuracy from the inspiral
waveform of binary neutron-star merger events. The colored
lines in Fig.~\ref{fig:Lambda} show that larger bounds on the maximum
mass push $\tilde\Lambda_{\rm min}$ to higher values [the coefficients
  $a,b,c$ used in Eq.~\eqref{eq:Lambda} are listed in
  Table~\ref{tab:table1}], while the upper bound
of~\citet{Altiparmak:2022} remains unaffected: $\tilde\Lambda_{\rm
  max}=-20+1800\,\mathcal{M}_{\rm chirp}^5$. The bottom part of
Fig.~\ref{fig:Lambda} shows again PDF slices for the chirp mass of
GW170817, which has been measured very accurately to be $\mathcal{M}_{\rm
  chirp}=1.188^{+0.004}_{-0.002}$. Combining the lower limits from all
the mass bounds allows us to set the following range for the lower bound
on the binary tidal deformability of GW170817 to be $\tilde\Lambda^{\rm
  min}_{1.186}\in[236,301]$ for $M_{\rm TOV} \in [2.18,
  2.52]\,M_{\odot}$. 

\begin{table}
 \centering
 \caption{Best-fit coefficients $a,b,c$ of Eq.~\eqref{eq:Lambda} for the
   lower bound of the binary tidal deformability parameter
   $\tilde{\Lambda}_{\rm min}$.}
 \label{tab:table1}
 \begin{tabular}{cccc} 
  \hline
  $M_{\rm TOV}~[M_\odot]$ & $a$ & $b$ & $c$\\
  \hline
  $\geq 2.00$  & $-50$ & $600$ & $4.7$ \\
  $\geq 2.18$  & $-45$ & $650$ & $4.6$\\
  $\geq 2.35$  & $-40$ & $750$ & $4.5$\\
  $\geq 2.52$  & $-20$ & $800$ & $4.4$\\
  \hline
 \end{tabular}
\end{table}

\begin{figure*}
    \centering
    \includegraphics[width=1\textwidth]{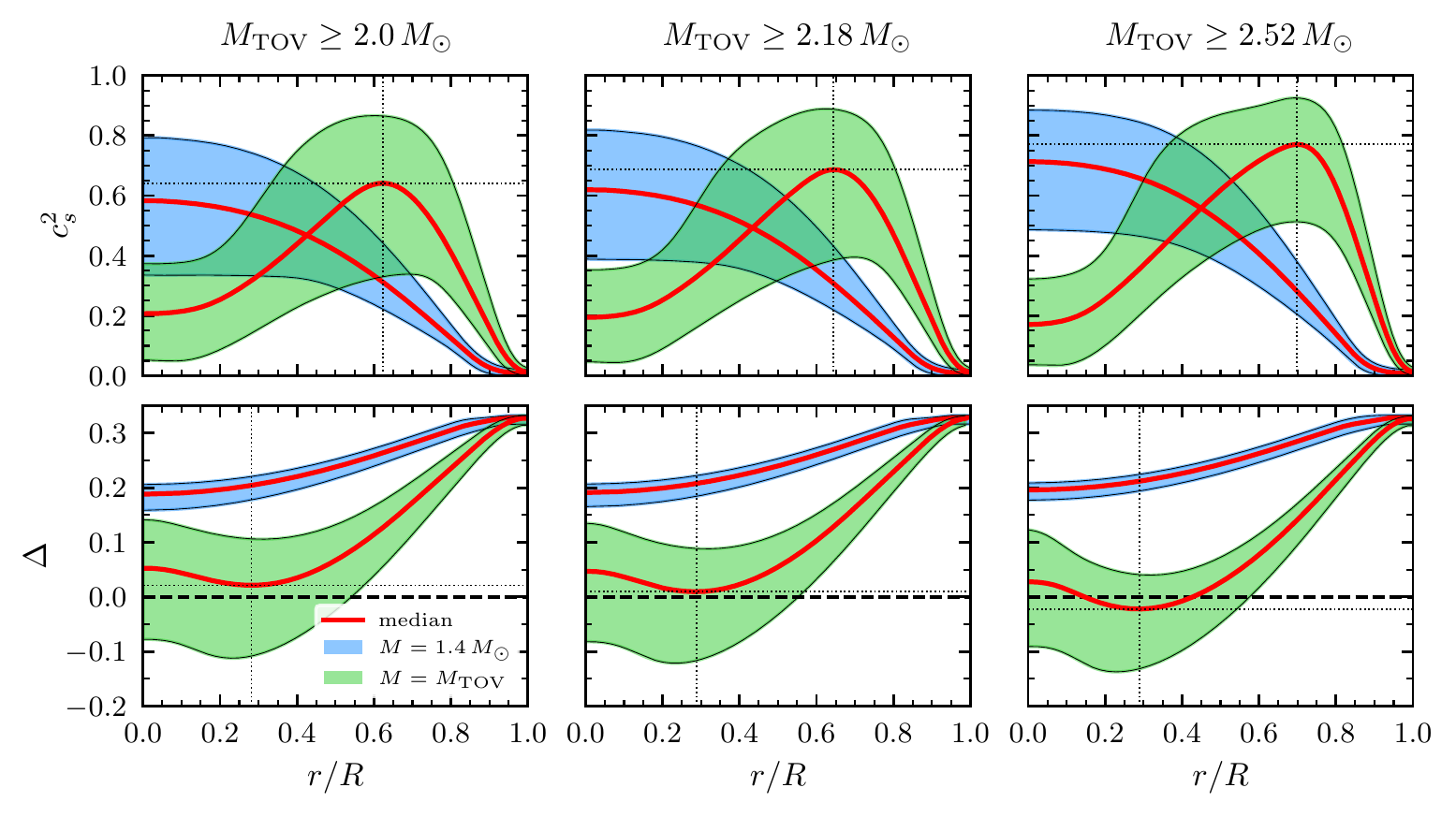}
    \caption{\textit{Top Row:} sound speed as function of normalized
      radial coordinate $r/R$ inside stars for different values of the
      mass constraint $M_{\rm TOV}\geq2.0-2.52\,M_\odot$ (left to right).
      Blue and green areas are $95\%$-confidence intervals for typical
      ($M=1.4\,M_\odot$) and maximally massive ($M=M_{\rm TOV}$) stars.
      Dotted black lines indicate the local sound speed maximum in
      maximally massive stars. \textit{Bottom Row:} The same as in the
      top row but for the conformal anomaly. Black dashed lines mark
      $\Delta=0$, while black dotted lines indicate the local minimum of
      the conformal anomaly.}
    \label{fig:insideStar}
\end{figure*}

We next move on to assessing how the new high-mass bounds on $M_{\rm TOV}$
affect the properties of the spatial distribution of the sound speed in
the stellar interior. We do this following a recent work of
ours~\citep{Ecker:2022}, where we have introduced a novel,
scale-independent description of the sound speed in neutron stars where
the latter is expressed in a unit-cube spanning the normalised radius,
$r/R$, and the mass normalized to the maximum one, $M/M_{\rm TOV}$. As
shown by~\citet{Ecker:2022}, a number of interesting results can be
deduced from this generic representation. In particular, the top-row of
panels in Fig.~\ref{fig:insideStar} shows the normalized radial
dependence of the sound speed for $M_{\rm TOV}\geq 2.0-2.52\,M_\odot$
(left to right). Blue and green-shaded areas denote the $95\%$-confidence
intervals for typical ($M=1.4\,M_\odot$) and maximally massive ($M=M_{\rm
  TOV}$) stars, respectively, where the red lines are instead used to
mark the median values (see Table~\ref{tab:table2} in
Appendix~\ref{sec:appendix}).

The finding by~\citet{Ecker:2022}, that light stars have monotonic
sound-speed radial profiles and heavy ones feature a local maximum in the
outer layers is even more pronounced when imposing larger bounds on
$M_{\rm TOV}$. Indeed, the larger the maximum-mass constraint, the more
asymmetric is the radial distribution of the PDFs, pushing the local maxima
of the medians to increasingly larger normalized radii. Exploiting this
behaviour, we can set bounds on the value and the location of the median
value of the sound-speed maximum in the interior of heavy stars
\begin{equation}
    c_{s,\rm max}^2\in [0.64,0.77]\,,\quad {\rm for} \quad r/R\in [0.62,0.72]\,.
\end{equation}
Clearly, for larger bounds on $M_{\rm TOV}$ the peak in the sound speed
develops already in lighter stars. For the strictest bound, \ie $M_{\rm
  TOV}\geq~2.52\,M_\odot$, this happens already in stars as light as
$\approx0.6\,M_{\rm TOV}$, whereas for $M_{\rm TOV}\geq~2.0\,M_\odot$ the
peak appeares only at $\approx 0.7\,M_{\rm TOV}$. It is important to
recall that the relevance of these results -- and their practical
application -- is that they can be included in nuclear-theory
calculations of modern EOSs to constrain the behaviour of the sound speed
in those regions where nuclear-theory predictions have large
uncertainties.

The bottom-row of Fig.~\ref{fig:insideStar} shows instead for
the first time the radial dependence of the conformal anomaly $\Delta$
given by Eq.~\eqref{eq:Delta}. Note how, in analogy with what happens for
the sound speed, stricter maximum-mass constraints lead to narrower
confidence intervals. Towards the stellar surface ($r/R=1$) the anomaly
approaches its vacuum value from below $\Delta\to 1/3$ independently of
the maximum-mass constraint, while the value of $\Delta$ in the stellar
center ($r/R=0$) does depend on $M_{\rm TOV}$. Note that in typical
neutron stars with masses $M\sim1.4\,M_\odot$, the anomaly decreases
monotonically from the surface ($\Delta=0$) towards the center, where it
reaches a median value of $\Delta\approx 0.2$. In turn, this implies
$\Delta\geq0$ inside such stars and that conformal symmetry is broken
everywhere in their interior. As shown in
Fig.~\ref{fig:insideStar}, the behaviour of the conformal anomaly in
light stars is very robust and depends only very weakly on the different
bounds set for the maximum mass (blue-shaded regions). This behaviour,
however, ceases to be true for maximally massive stars (green-shaded
regions) where the $95\%$-confidence variance is far larger and where the
conformal anomaly exhibits a local minimum $\Delta_{\rm min}\approx 0$ at
$r/R\approx0.3$. Furthermore, depending on the constraint on $M_{\rm
  TOV}$ the median value of $\Delta$ (red lines) can either be positive
or negative (or both) in the innermost regions of the star. More
specifically, for $M_{\rm TOV}\geq 2.0\,M_\odot$ we find $\Delta_{\rm
  min}>0$, while for $M_{\rm TOV}\geq 2.52\,M_\odot$ the anomaly is
$\Delta_{\rm min}<0$, somewhat in contrast with the expectations
of~\citet{Fujimoto:2022}. In the range $M_{\rm TOV}\geq
2.18-2.35\,M_\odot$ the value of $\Delta_{\rm min}$ is very close to
zero. In other words, depending on the bound set on $M_{\rm TOV}$,
maximally massive stars can either have zero, one, or two layers where
conformal symmetry ($\Delta=0$) is restored.

\begin{figure}
    \centering
    \includegraphics[width=0.475\textwidth]{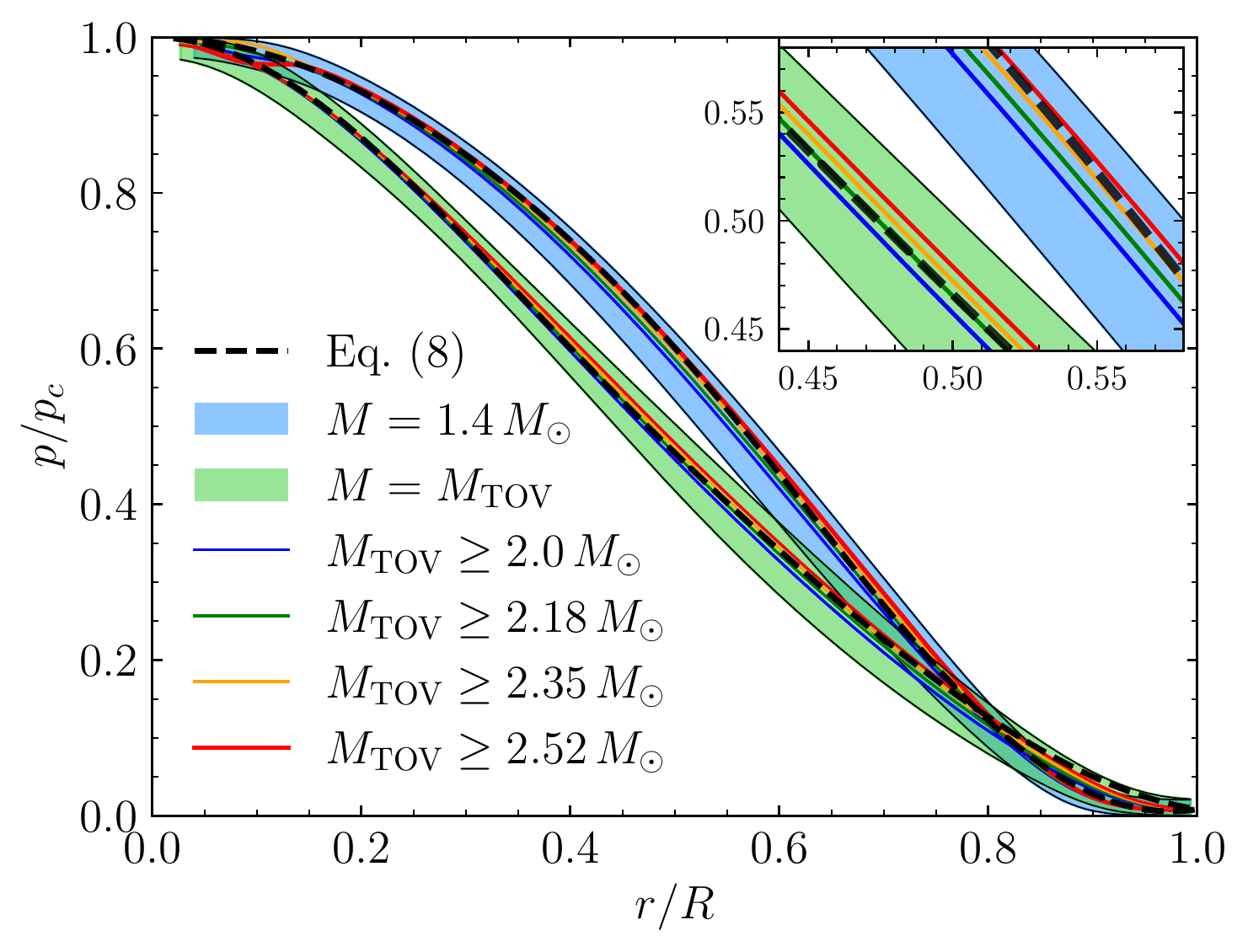}
    \caption{Quasi-universal radial distribution of the pressure in the
      stellar interior $p(r/R)/p_c$. As in previous figures, blue and
      green-shaded areas represent the $95\%$-confidence intervals for
      typical ($M=1.4\,M_\odot$) and maximally massive ($M=M_{\rm TOV}$)
      stars, respectively. Coloured lines are the medians for different
      lower bounds on $M_{\rm TOV}$, while the black dashed lines report
      their best fits using Eq.~\eqref{eq:fit_px}.}
    \label{fig:pUniv}
\end{figure}

Finally, we use Fig.~\ref{fig:pUniv} to present what is a particularly
interesting result: a novel quasi-universal law for the radial
distribution of the pressure inside neutrons stars. Adopting the same
convention for the shading of the $95\%$-confidence areas and of the
medians of the PDFs, Fig.~\ref{fig:pUniv} reports the (normalized) radial
dependence of the pressure $p(r/R)/p_c$, where $p_c:=p(0)$ is the central
pressure. Note how the variance of the $\approx 10^6$ EOSs considered is
extremely small both in the case of reference neutron stars with
$1.4\,M_{\odot}$ (blue-shaded area) and for the maximally massive stars
(green-shaded areas), giving a variation from the median value of the
median that is $\lesssim 8\%$ (see inset). Also quite remarkable is how
the functional behaviour is essentially insensitive to the high-mass
bound on $M_{\rm TOV}$, with differences that are smaller than a few
percent even when the largest constraint is imposed on the maximum mass.
As a result, this novel relation will likely remain unaffected by more
accurate and future maximum-mass measurements and therefore already now
represents a robust prediction for the pressure distribution in the
interior of neutron stars.

The very tight and quasi-universal behaviour of the pressure profile can
be accurately described in terms of a simple fitting formula of the type
\begin{equation}
  \label{eq:fit_px}
  \frac{p(x)}{p_c}=e^{\alpha x^2}+e^{\alpha}\frac{\cos \beta x-1}{\cos \beta-1}\,,
\end{equation}
with $x:=r/R$ and where the coefficients for maximally massive stars (or
for typical $1.4\,,M_\odot$ stars) are given by
$\alpha=6~(6.5)\,,\beta=-5.5~(-5.5)$, respectively. Since expression
\eqref{eq:fit_px} refers to EOSs that by construction satisfy all of the
astronomical and QCD constraints, it can be used as a ``sanity-check'' in
the construction of EOSs that start instead from basic principles of
theoretical nuclear physics.

\section{Conclusion}

Motivated by the recently announced measurement by~\citet{Romani:2022jhd}
of the neutron-star mass in PSR J0952-0607 with $M=2.35\pm0.17~M_\odot$,
we have studied the impact that large bounds on the maximum mass have on
the EOS of nuclear matter and on neutron-star properties. To this scope,
we have employed an agnostic approach for the construction of EOSs
based on a sound-speed parameterisation and that we have carefully tested
and exploited in previous works~\citep{Altiparmak:2022, Ecker:2022}.

In this way, we have found that increasingly large bounds on the maximum
mass $M_{\rm TOV}$ do change the statistical properties of the EOSs and
the corresponding neutron-star characteristics. In particular, the
largest bound on the maximum mass coming from PSR J0952-0607 decreases
the EOS uncertainty at neutron-star densities significantly, squeezing
the $95\%$-confidence interval for the pressure's PDF to a narrow band
around $p \approx 200~{\rm MeV}/{\rm fm}^3$ at energy densities $e\approx
600~{\rm MeV/fm}^3$. Furthermore, raising the maximum-mass bound from
$2.0\,M_\odot$ to $2.52\,M_\odot$ increases systematically the radius of
a typical neutron star with $1.4\,M_{\odot}$, taking it from a median
value $R_{1.4}=12.48^{+0.75}_{-1.14}\,\rm km$ over to
$R_{1.4}=12.97^{+0.28}_{-0.64}\,\rm km$, reducing at the same time the
$95\%$-confidence level by almost $50\%$. This behaviour is rather
natural and reflects the fact that larger maximum masses require stiffer
EOSs and, in turn, stellar models that have on average larger radii.
This effect is even more pronounced in the most massive stars, where we
find an increase in radius by almost one kilometre for a reference star
with a mass of $2.0\,M_\odot$. Because the measurements of the NICER
experiment can be used to set lower bounds on the stellar radii, having
very large maximum masses has the drawback of making NICER's constraints
largely ineffective.

Another important quantity in our analysis is the (analytic) relation
between the binary tidal deformability and the chirp mass
$\tilde\Lambda(\mathcal{M}_{\rm chirp})$, which has the potential of
relating directly a quantity measured with great precision in
gravitational-wave detections $\mathcal{M}_{\rm chirp}$ with a sensitive
property of the EOS $\tilde\Lambda$. We have therefore explored what is
the impact that larger bounds on the maximum mass have on the resulting
relations and found that while the upper bound $\tilde\Lambda_{\rm max}$
is essentially insensitive to changes on $M_{\rm TOV}$, the lower bound
$\tilde\Lambda_{\rm max}$ increases systematically with larger bounds. 

Our analysis has also allowed us to explore the scale-independent radial
behaviour of the sound speed and, for the first time, of the conformal
anomaly $\Delta$ within the stellar interior. In this way, we have found
that increasing the mass bound from $2.0\,M_\odot$ to $2.52\,M_\odot$
pushes the maximum sound speed in maximally massive stars from $c_{s,\rm
  max}^2\simeq0.64$ up to $c_{s,\rm max}^2\simeq0.77$ 
 and further away from the neutron-star center, \ie from
$r/R\simeq0.62$ to $r/R\simeq0.72$.

We also presented the first results for the radial dependence of the
conformal anomaly $\Delta$ in the neutron-star interior inspired by the
question formulated by~\citep{Fujimoto:2022} on whether $\Delta$ is
positive definite. Our findings indicate that indeed $\Delta > 0$ in the
interior of typical stars with masses of $1.4\,M_\odot$, for which
$\Delta$ decreases monotonically from its vacuum value $\Delta=1/3$ at
the stellar surface down to $\Delta\approx 0.2$ at the neutron-star
center. However, for maximally massive stars, the conformal anomaly
exhibits a nonmonotonic behaviour inside the star and we show that the
value of the anomaly at the local minimum can be either positive, zero or
negative, depending on the value for lower bound on $M_{\rm TOV}$.

Finally, we have reported a novel quasi-universal law for the radial
distribution of the pressure inside neutrons stars when cast into a
scale-independent manner. The variance around the median is surprisingly
small ($\lesssim 8\%$) and the functional behaviour is essentially
insensitive to the high-mass bound on $M_{\rm TOV}$, with differences
that are smaller than a few percent even when the largest constraint is
imposed on the maximum mass. As a result, this relation will likely
remain unaffected by more accurate future maximum-mass measurements
and can be used as a ``sanity-check'' in the theoretical construction of
EOSs.

There exists a number of possibilities to generalize our work.
One such possibility is to explicitly include first and/or second 
order phase transitions in the EOS construction.
While our method in principle includes such EOSs, they are statistically 
underrepresented in our ensemble, rendering their impact negligible 
on the final result.
It would also be interesting to test the universal relation
for the pressure profile presented in this work in other gravity theories than 
General Relativity.
Finally, another interesting possibility would be to generalize our statistic analysis
to rapidly spinning stars, which is numericlally more demanding, but feasible 
and we plan to perform such simulations in future work.

\section*{Acknowledgements}

We thank T.~Gorda and A.~Kurkela for useful discussions. Partial funding
comes from the State of Hesse within the Research Cluster ELEMENTS
(Project ID 500/10.006), by the ERC Advanced Grant ``JETSET: Launching,
propagation and emission of relativistic jets from binary mergers and
across mass scales'' (Grant No. 884631). CE acknowledges support by the
Deutsche Forschungsgemeinschaft (DFG, German Research Foundation) through
the CRC-TR 211 ``Strong-interaction matter under extreme conditions''--
project number 315477589 -- TRR 211. LR acknowledges the Walter Greiner
Gesellschaft zur F\"orderung der physikalischen Grundlagenforschung
e.V. through the Carl W. Fueck Laureatus Chair. The calculations were
performed on the local ITP Supercomputing Clusters Iboga and Calea.

\section*{Data Availability}
Data is available upon reasonable request from the corresponding
author.

\bibliographystyle{mnras}
\bibliography{aeireferences.bib} 



\appendix

\section{Impact of QCD constraints}
\label{sec:appendix}

We first collect in Table~\ref{tab:table2} the numerical values of the
various EOSs and neutron-star properties discussed in the main text. We
list results for cases where we impose constraints that are only
``astro'', or only ``QCD'', or their combination ``astro+QCD''. The QCD
constraints have the largest impact on the minimum values of the
conformal anomaly $\Delta_{\rm min}$ and the sound speed in the neutron
star center $c_{s,\rm center}^2$. Note that there is a clear trend,
namely that $\Delta_{\rm min}$ is systematically underestimated, while
$c_{s,\rm center}^2$ is systematically over estimated if QCD is not
imposed.

\begin{table*}
 \centering
 \caption{Impact of constraints on neutron-star properties. Taking as a
   reference maximally massive stars of mass $M=M_{\rm TOV}$, the various
   columns report: the minimum values of the conformal anomaly
   $\Delta_{\rm min}$, the sound speed at the neutron-star center
   $c_{s,\rm center}^2$, the maximum sound speed $c_{s,\rm max}^2$ and
   its radial location inside stars ${r_{\rm max}}/{R}$. Also listed are
   the radii $R_{1.4}$ ($R_{2.0}$) of stars with mass
   $M=1.4\,(2.0)\,M_\odot$, the binary tidal deformability
   $\tilde\Lambda_{1.186}$ of a GW170817-like event and the minimum
   compactness $\mathcal{C}_{\rm TOV}^{\rm min}$ of maximally massive
   stars. Quantities with error estimates correspond to median values and
   uncertainties to $95\%$-confidence intervals. Note that are reported
   values in which either the ``astro'' or the ``QCD'' or both
   constraints ``astro+QCD'' are imposed.
 }
 \label{tab:table2}
\renewcommand{\arraystretch}{1.5}
 \begin{tabular}{cccccccccc} 
   \hline
  $M_{\rm TOV}/M_\odot$ & Constraints &
  $\Delta_{\rm min}$ & $c^2_{s,\rm center}$ & $c^2_{s,\rm max}$ & ${r_{\rm max}}/{R}$ & $R_{1.4}~[{\rm km}]$ & $R_{2.0}~[{\rm km}]$ & $\tilde\Lambda_{1.186}$ & $\mathcal{C}_{\rm TOV}^{\rm min}$\\
  \hline
	$\geq 2.00$ & astro     & $-0.05^{+0.14}_{-0.18}$  & $0.38^{+0.40}_{-0.33}$  & $0.68^{+0.21}_{-0.34}$ & $0.63$ & $12.59^{+0.65}_{-1.23}$ & $12.49^{+1.33}_{-1.63}$ & $415^{+283}_{-178}$ & $0.222$\\
              & astro+QCD & $+0.02^{+0.08}_{-0.13}$ & $0.21^{+0.17}_{-0.15}$ & $0.64^{+0.22}_{-0.31}$ & $0.62$ & $12.48^{+0.75}_{-1.14}$ & $12.32^{+1.43}_{-1.47}$ & $412^{+282}_{-176}$ & $0.221$ \\
  \hline
	$\geq 2.18$ & astro     & $-0.06^{+0.13}_{-0.17}$  & $0.39^{+0.38}_{-0.33}$  & $0.73^{+0.19}_{-0.34}$ & $0.65$ & $12.72^{+0.54}_{-1.13}$ & $13.00^{+0.86}_{-1.61}$ & $434^{+266}_{-180}$ & $0.236$ \\
              & astro+QCD & $+0.01^{+0.08}_{-0.13}$ & $0.19^{+0.16}_{-0.15}$ & $0.69^{+0.20}_{-0.30}$ & $0.64$ & $12.67^{+0.67}_{-1.05}$ & $12.92^{+0.89}_{-1.42}$ & $449^{+253}_{-183}$ & $0.235$ \\
  \hline
	$\geq 2.35$ & astro     & $-0.08^{+0.13}_{-0.16}$ & $0.42^{+0.37}_{-0.35}$ & $0.74^{+0.18}_{-0.31}$ & $0.68$ & $12.82^{+0.43}_{-0.92}$ & $13.20^{+0.66}_{-1.21}$ & $524^{+205}_{-223}$ & $0.251$ \\
              & astro+QCD & $+0.00^{+0.07}_{-0.12}$ & $0.18^{+0.15}_{-0.14}$ & $0.72^{+0.18}_{-0.28}$ & $0.68$ & $12.85^{+0.40}_{-0.83}$ & $13.25^{+0.60}_{-1.09}$ & $485^{+231}_{-191}$ & $0.250$ \\
  \hline
	$\geq 2.52$ & astro     & $-0.10^{+0.12}_{-0.15}$ & $0.44^{+0.38}_{-0.37}$  & $0.77^{+0.16}_{-0.28}$ & $0.70$ & $12.94^{+0.34}_{-0.76}$ & $13.41^{+0.47}_{-0.94}$ & $498^{+303}_{-215}$ & $0.266$ \\
              & astro+QCD & $-0.02^{+0.06}_{-0.11}$ & $0.17^{+0.15}_{-0.13}$  & $0.77^{+0.15}_{-0.26}$ & $0.70$ & $12.97^{+0.28}_{-0.64}$ & $13.47^{+0.42}_{-0.80}$ & $517^{+206}_{-216}$ & $0.266$ \\
  \hline
	      & QCD       & $+0.02^{+0.12}_{-0.15}$  & $0.25^{+0.30}_{-0.19} $ & $0.54^{+0.28}_{-0.31}$  & $0.55$ & $11.02^{+3.24}_{-3.20}$ & $13.00^{+2.30}_{-2.54}$ & $285^{+598}_{-256}$& $-$\\
  \hline
 \end{tabular}
\end{table*}

\begin{figure}
    \centering
    \includegraphics[width=0.475\textwidth]{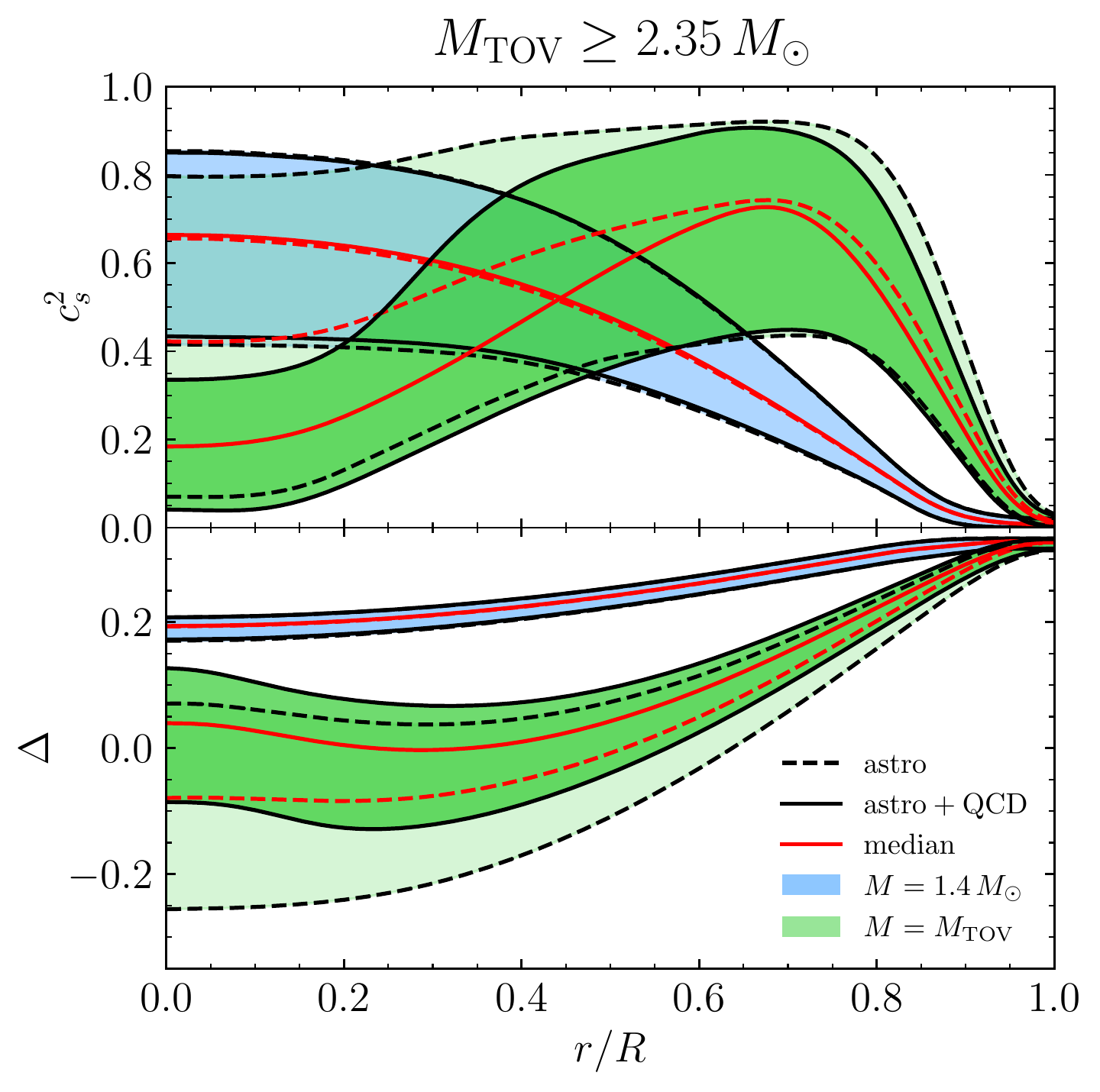}
	\caption{The same as in Fig.~\ref{fig:insideStar} but when
          consider the differential impact of the astro (black solid
          lines) and astro+QCD (black dashed lines) constraints. Red
          lines mark the median, with solid (dashed) lines including
          (exclude) the QCD constraints.}
    \label{fig:pQCDimpact}
\end{figure}

Next, we consider how perturbative QCD boundary conditions imposed in
model-agnostic approaches can have a relevant impact on the EOSs at
densities realised in neutron stars close to their maximum
mass~\citep{Somasundaram:2022,Gorda:2022}. We study the impact of these
boundary conditions by comparing results where they are imposed to those
where they are not imposed. While the impact on our estimated for
neutron-star radii is rather small, the relative differences of the
conformal anomaly and the sound speed inside maximally massive stars can
be significant, which allows us to identify these quantities as
particular sensitive to the perturbative QCD boundary conditions. In
Figure~\ref{fig:pQCDimpact} we exemplify the impact of QCD constraints on
the sound-speed distribution and the conformal anomaly inside stars for
the case $M_{\rm TOV}\geq 2.35\,M_\odot$.

We find the impact of QCD on typical stars is negligible, but this is
entirely different in maximally massive stars. There the quantitative and
even the qualitative behavior of $c_s^2$ and $\Delta$ in the core is very
different. Without imposing QCD, the conformal anomaly decreases
monotonically towards the stellar center, where its value is negative.
Also the confidence interval gets very wide, which is a sign that without
the QCD boundary conditions $\Delta$ and $c_s^2$ are not well constrained
in the centres of heavy stars. This can be seen particularly well on
$c_s^2$, whose confidence interval increases form a quite narrow band
$\approx 0-0.3$ in the constrained case to $\approx 0-0.8$ in the
unconstrained case. In consequence, there is $100\%$ difference in the
median values of the sound speed $c^2_{s,\rm center}=0.2$ (with QCD)
versus $0.4$ (without QCD). In summary, QCD has negligible impact on the
mass-radius relation, but is important to determine the radial
distribution of the sound speed and the conformal anomaly of massive
stars.


\bsp	
\label{lastpage}
\end{document}